\documentclass[amssymb,amsmath,aip,groupedaddress,pra]{revtex4-2}

\usepackage[english]{babel}
\usepackage{graphicx}
\usepackage{color}
\usepackage{bm}
\usepackage{amsmath}
\usepackage{amssymb}

\DeclareGraphicsExtensions{.eps,.pdf,.png,.jpg}

\begin{document}

\title{Stationary striations in plasma, created by a short microwave pulse in a waveguide filled with a neutral gas }

\author{Y. Bliokh}
\affiliation{Physics Department, Technion, Israel Institute of Technology, Haifa 320003, Israel}
\author{Y. Cao}
\thanks{Corr.author}
\email{neo.cao.yang@gmail.com}
\affiliation{Physics Department, Technion, Israel Institute of Technology, Haifa 320003, Israel}
\author{V. Maksimov}
\affiliation{Physics Department, Technion, Israel Institute of Technology, Haifa 320003, Israel}
\author{A. Haim }
\affiliation{Physics Department, Technion, Israel Institute of Technology, Haifa 320003, Israel}
\author{J.G. Leopold }
\affiliation{Physics Department, Technion, Israel Institute of Technology, Haifa 320003, Israel}
\author{A. Kostinsky}
\affiliation{Physics Department, Technion, Israel Institute of Technology, Haifa 320003, Israel}
\author{Ya.E. Krasik}
\affiliation{Physics Department, Technion, Israel Institute of Technology, Haifa 320003, Israel}

\begin{abstract}

It was observed experimentally that after crossing a waveguide filled with a neutral gas, a short powerful microwave pulse leaves a periodic glow of plasma along the waveguide, persisting several tens of nanoseconds. A theoretical model is presented which in combination with numerical simulations proposes a possible explanation of this novel phenomenon.

\end{abstract}

\maketitle

\section{Introduction}

Interaction of strong electromagnetic waves with plasma has always attracted considerable attention. Charged particles acceleration in the wake wave exited by a powerful laser pulse propagating in plasma \cite{Tajima-1979} is at present of particular interest because of significant experimental and theoretical achievements and practical importance (see, e.g., \cite{Esarey-2009, Hooker-2013, Tajima-2020}).

The non-linear interaction of electromagnetic pulses with plasma when the microwave and plasma frequencies are of the same order of magnitude, remains permanently in researchers' field of view. Recent progress in generating extremely intense, hundreds of MW, sub-nanosecond microwave pulses \cite{Eltchaninov-2004, Rostov-2016} allows one to observe experimentally and study such phenomena as ionization-induced self-channeling of a microwave beam,\cite{Shafir-2018}, frequency shift in the wake excited in a plasma-filled waveguide\cite{Cao-2020}, complete absorption of a microwave pulse in plasma,\cite{Cao-2021} frequency up-shift and pulse compression in a propagating self-generated ionization front \cite{Cao-2023}. 

\begin{figure}[tbh]
	\centering \scalebox{0.15}{\includegraphics{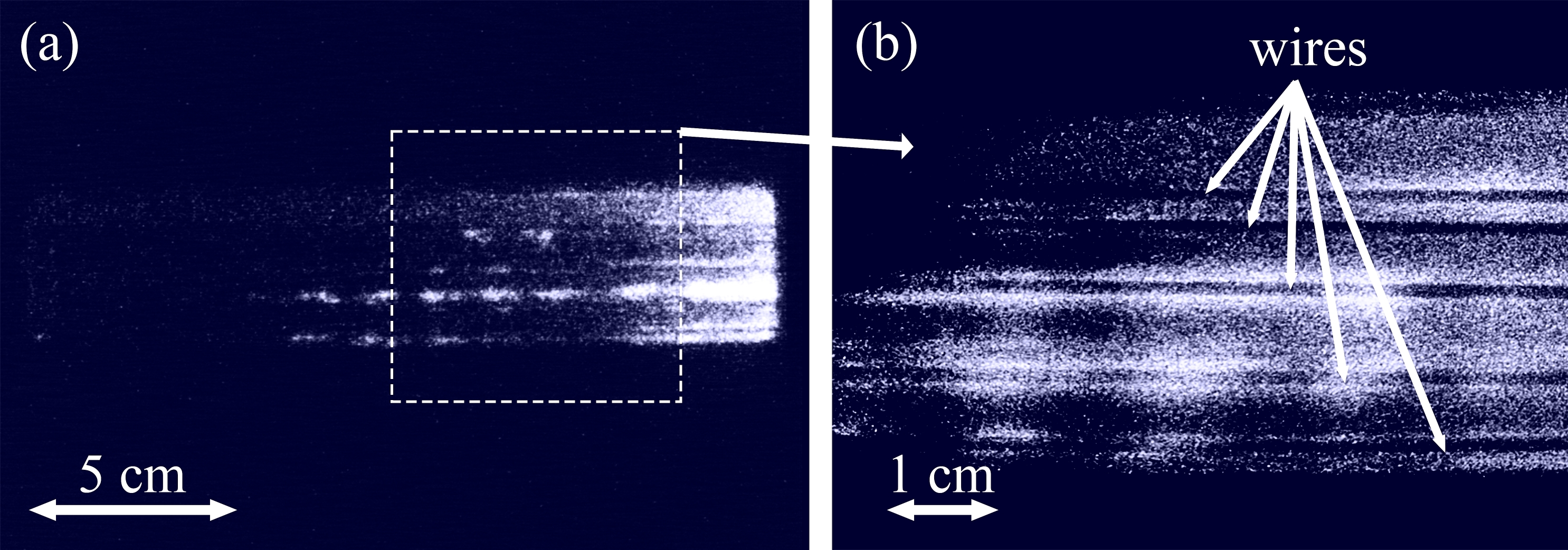}}
	\caption{Plasma glow at 18 torr helium pressure: (a) broad field-of-view and (b) detailed close-up view. Frame duration of 1.5 ns. The time delay of 12 ns with respect to the microwave pulse entering the gas chamber. }
	\label{Strat}
\end{figure}

In this paper a new phenomenon is described which accompanies a neutral gas ionization by intense sub-nanosecond microwave pulse propagating in a waveguide. It is suggested that a periodic stratification of the plasma forms near the waveguide wall and its glow is observed (see Fig.~\ref{Strat}), for helium within a pressure range of 7-20 torr and for air within 1-3 torr. The plasma glow appears approximately 5 ns after the microwave pulse enters the gas tube and persists for ~70 ns.
 
There is nothing unexpected in that perturbations remain in the plasma after the electromagnetic pulse's passage. The wake excited by a short laser pulse is an example of such perturbation which  has received most of the attention. This perturbation propagates together with the pulse and behind it, that is, in the\textit{ pulse}'s frame of reference, it does not move. The periodic stratification demonstrated in Fig.~\ref{Strat}, is static in the\textit{ laboratory} frame of reference and appears in the same positions along the waveguide from shot to shot. The glow is concentrated near the wires, forming the waveguide boundary, that implies existence of the radial electric field, periodically distributed along the waveguide. Establishing the reasons of this field initiation and its long lifetime (about two orders of magnitude longer than the Langmuir oscillations period) is of interest to basic plasma physics. The purpose of the present article is to propose a physical model which explains this phenomenon.  

\section{Experimental setup}

\begin{figure}[tbh]
	\centering \scalebox{0.6}{\includegraphics{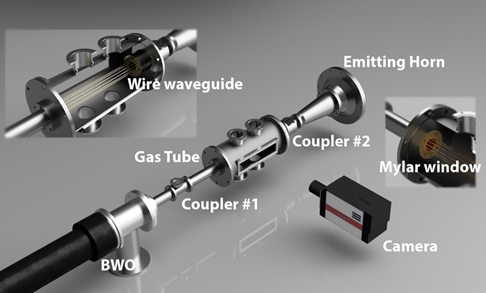}}
	\caption{Experimental setup}
	\label{Setup}
\end{figure}

The experimental setup (see Fig.~\ref{Setup}) used in this research is similar to that described in \cite{Cao-2023}. 
The high-power microwave (HPM) pulse ($\sim 250 - 300$ MW, $\sim 0.35$ ns at Full Width Half Maximum (FWHM), $\sim 9.6$ GHz, $TM_{01}$ mode) is generated by a super-radiant backward wave oscillator (SR-BWO) (see \cite{Eltchaninov-2004, Rostov-2016} for details). The output of the SR-BWO is connected to a 2.8 cm diameter circular waveguide with a calibrated coupler (\#1) at its center. This waveguide is followed by the gas tube, filled with helium or air at the desired pressure. Inside the gas tube, {the solid-wall circular waveguide is replaced by twelve 1 mm diameter stainless steel wires parallel to the waveguide axis and distributed uniformly along the circumference of a circle of the same diameter as the corresponding solid wall waveguide. The use of such a ``transparent'' waveguide section allows access for diagnostic and optical observation of the processes, induced by the microwave pulse. Experimental measurements \cite{Cao-2019} and electromagnetic} simulations confirm that the wave propagating along this waveguide is practically indistinguishable from the wave propagating along a solid-wall waveguide. A second calibrated coupler (\#2) is connected at the exit from the gas tube followed by an impedance-matched horn antenna. Couplers \#1 and \#2 measure the incident, transmitted, and reflected wave forms acquired by an Agilent DSO81204B oscilloscope (12 GHz, 40 Gs/s). At the entrance and exit of the gas tube, 0.3 mm thick Mylar interface windows are installed to separate the vacuum/gas/vacuum media. A fast framing intensified 4QuikE ICCD camera (Stanford Computer Optics) operating with a frame duration of 1.5 ns was used to capture the light emission from the plasma formed by the HPM pulse.

The plasma light emission patterns are shown in Fig.~\ref{Strat}. One can see periodic plasma glow (striations) near wires which appears approximately 5 ns after the HPM pulse has left the wire waveguide and this glow is present for ~70 ns. {Black  stright lines, visible in Fig.~\ref{Strat}, are waveguide-forming wires, which are facing the observer side. It means that the glow appears at the inner,  presented to the axis sides of the wires.} Such striations were obtained in helium within the pressure range of 7-20 torr, and in air at pressures of 1 - 3 Torr, which corresponds to the transparency window of the HPM pulse in air. By capturing frames at different time delays between the HPM pulse registered by coupler \#1 and the 4QuilE camera frame, it can be stated that the striations are stationary and do not move along the path of the HPM pulse propagation. {Note, that the microwave wave length is about 6 cm in the empty waveguide and increases when the plasma density grows, while typical longitudinal period of the striations is much smaller, about 2 cm.}

\section{Ionization}

\begin{figure}[h] 
	\centering \scalebox{0.4}{\includegraphics{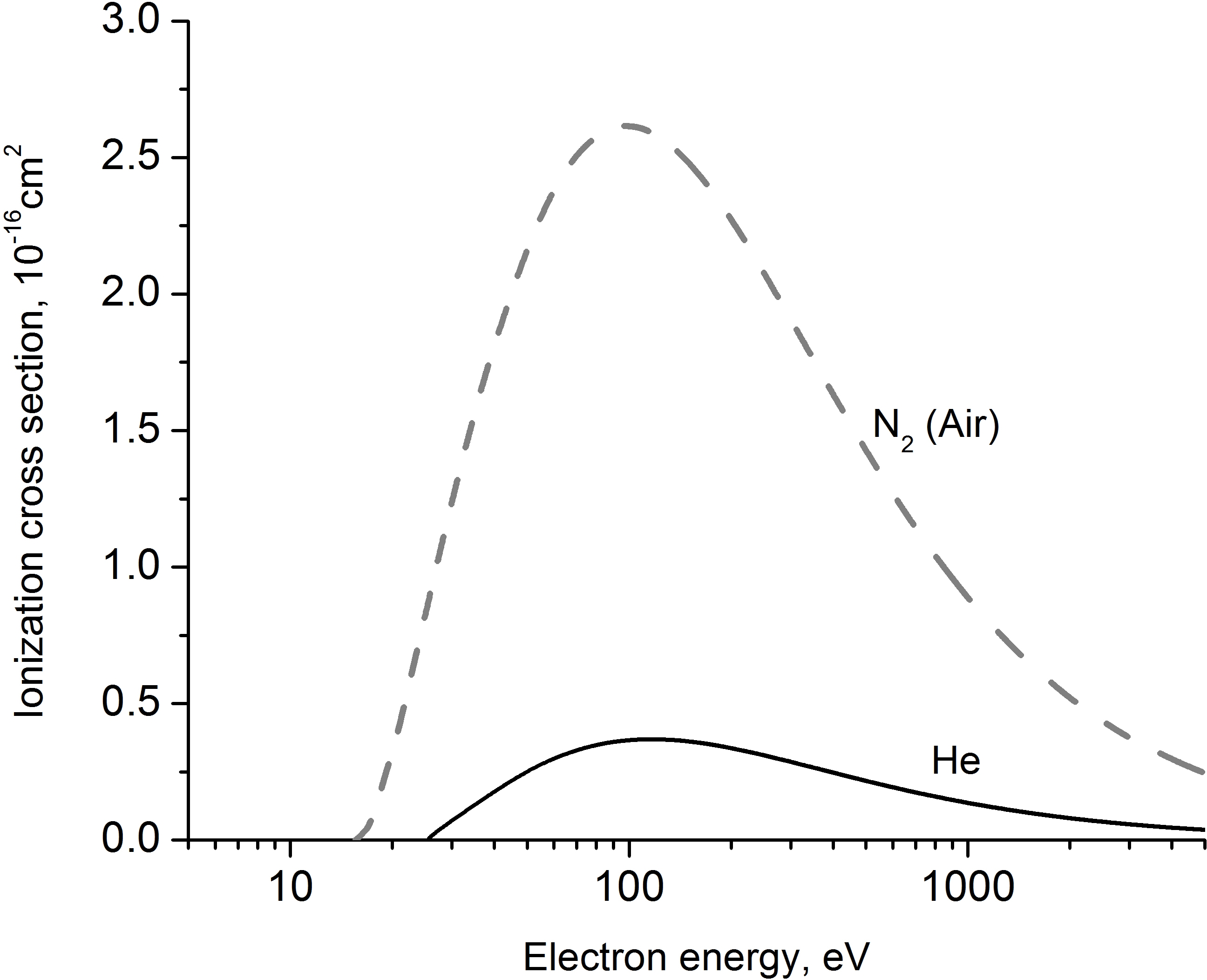}}
	\caption{{Electron impact ionization cross section $\sigma(w)$ as function of electron energy. Dashed line -- molecular nitrogen (air), solid line -- helium \cite{NIS}. } }
	\label{FigA}
\end{figure}

{If the wave frequency is not far from the cut-off frequency, as is the case in these experiments (operating frequency 9.6 GHz, cut-off frequency 8.19 GHz),} the electric field of a symmetric TM$_{01}$ microwave mode and the energy of electrons, oscillating in the wave fields, are maximal near the axis. When the microwave power is moderate (a few MW) the plasma is created mostly near and its density is maximal on the axis. When the microwave power is sufficiently high, the region of maximal ionization rate shifts from the axis toward the waveguide wall. The reason is a non-monotonic dependence of the electron impact ionization cross section $\sigma(w_e)$ on the electron energy $w_e$. The cross section is maximal when the electron energy is of the order of 100 eV and decreases with increasing electron energy. {As an example, the electron impact ionization cross section for helium and air are shown in Fig.~\ref{FigA}. For microwave power of the order of tens of MW or higher, the electron oscillating energy reaches several keV
near the axis, so that the ionization cross sections here is much smaller than its maximal value.} In contrast, near the waveguide wall, where the wave electric field is smaller than near the axis, the ionization cross section is close to its maximal value. 

\begin{figure}[h] 
	\centering \scalebox{0.4}{\includegraphics{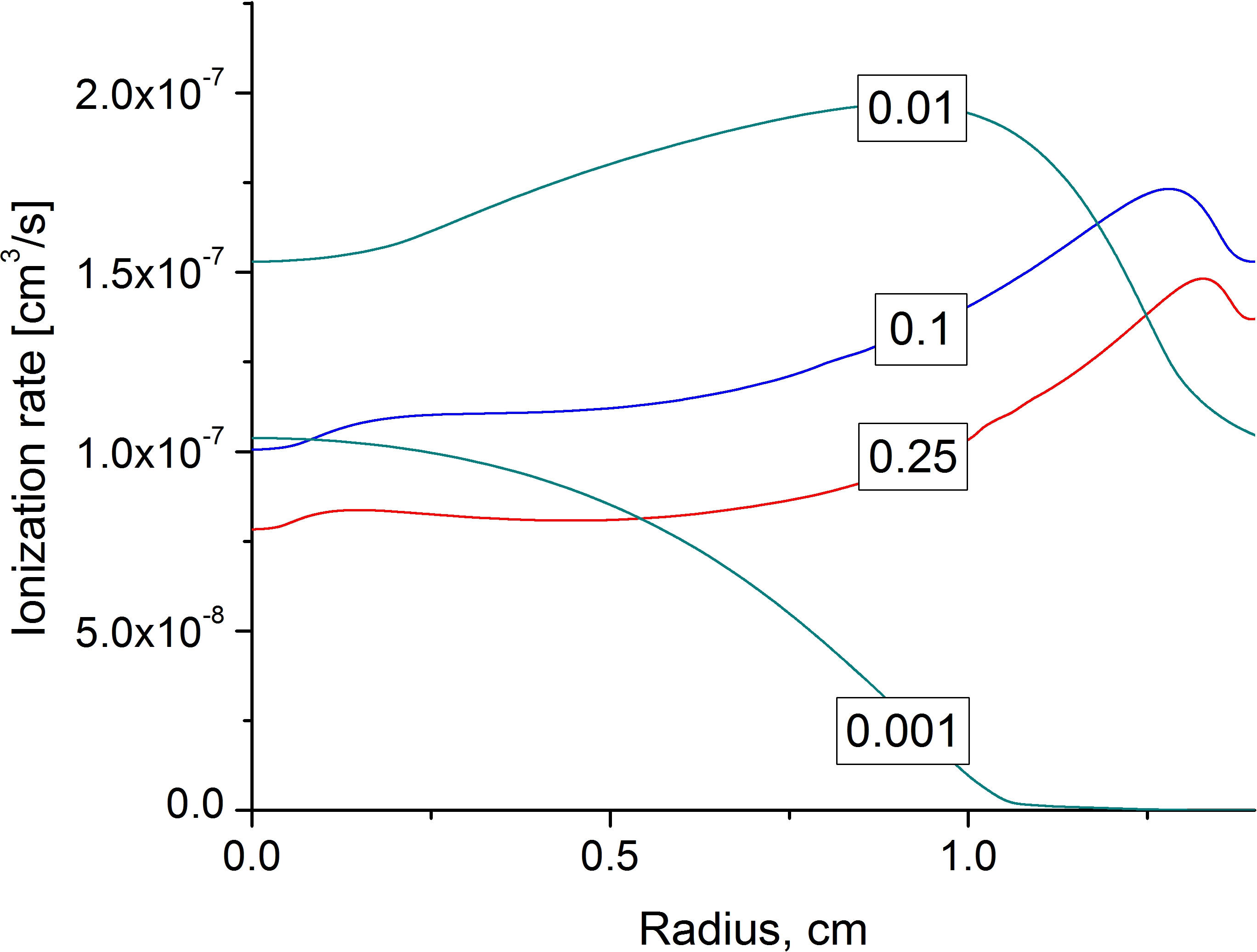}}
	\caption{Ionization rate $\langle{I}(r;P)\rangle$ for He as function of radius for various values of microwave power $P$ (in GW). }
	\label{Fig2}
\end{figure}

The time evolution of the electron density, $n_e(r,t)$, is described by the  equation
\begin{equation}
	\label{Ion_eq}
	\frac{d}{dt}\ln n_e(t)=n_g\sigma[w(r,t)]|v(r,t)|\equiv n_gI(r;P),
\end{equation}
where $n_g$ is the neutral gas density, $P$ is the microwave power, $v$ is the electron velocity, and $\sigma[w(r,t)]$ is the ionization cross section, which depends on the electron kinetic energy $w=mv^2/2$. If the duration of the electromagnetic pulse, $t_{\rm pulse}$, is larger than the field oscillation period, {it is convenient to use the ionization rate averaged over a period of the oscillations,   $\langle I(r,t)\rangle$.
The dependencies of the averaged ionization rate $\langle I(r;P)\rangle$ on the radius, which are shown in Fig.~\ref{Fig2} for various values of the microwave power $P$,  was calculated numerically using the cross section data and radial profile of the TM$_{01}$ mode electric fields.}  

\begin{figure}[h] 
 	\centering \scalebox{0.4}{\includegraphics{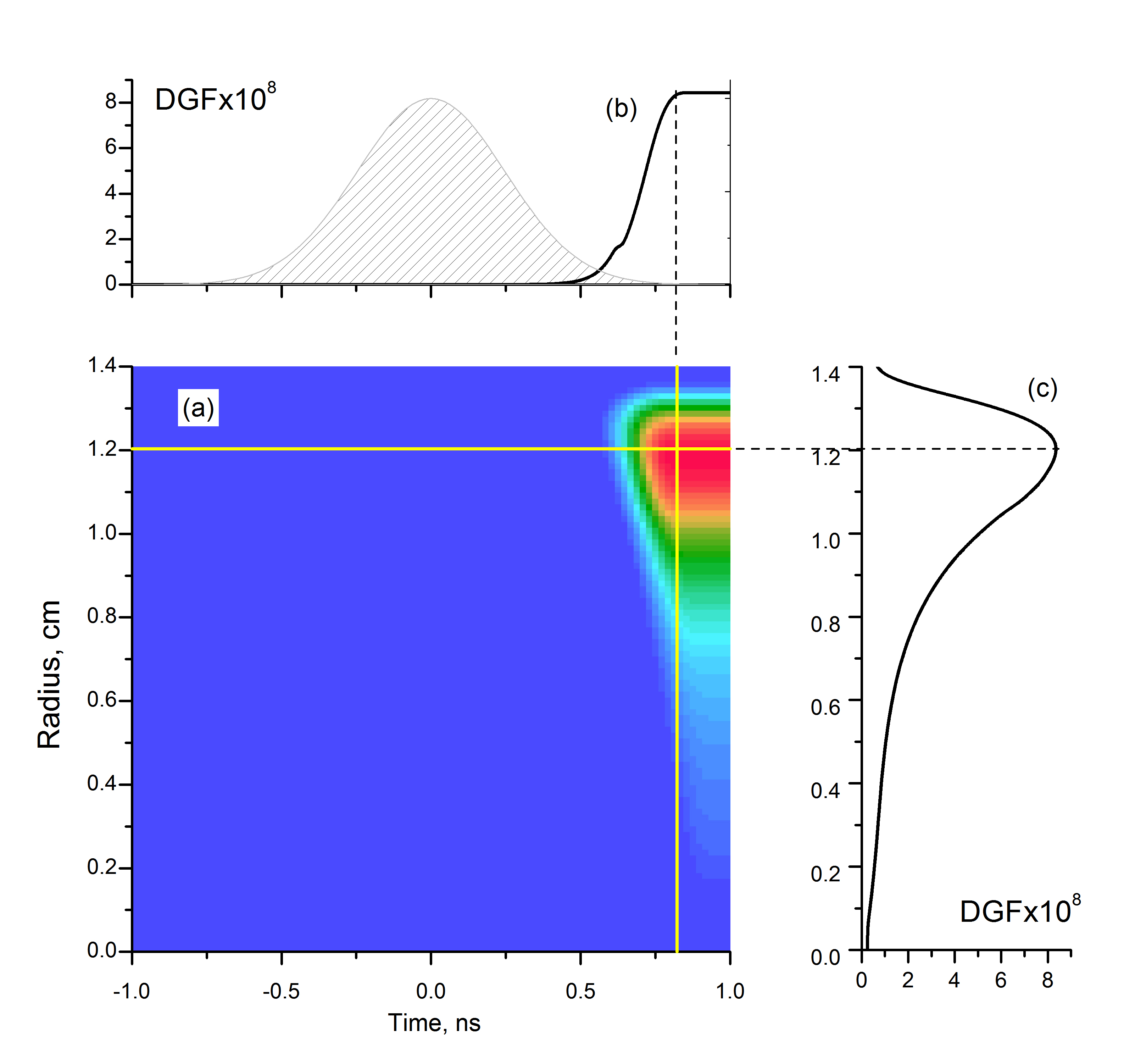}}
 	\caption{The temporal and radial evolution of the plasma density growth factor (DGF) $n_e(r,t)/n_0$. For helium, at 2.5 torr pressure, pulse power 300 MW, pulse duration 0.35 ns. The temporal profile of the normalized pulse power is shown as a shaded area in (b).}
 	\label{Fig3}
 \end{figure}

The radial profile of the plasma density, which remains in the waveguide behind the pulse, can be calculated as
\begin{equation}
	\label{eqA3}
	n_e(r,t)=n_0\exp\left[n_g\int_{-\infty}^t\langle{I}(r;P(t^\prime))\rangle dt^\prime\right],
\end{equation} 
where $n_0$ is the initial electron density, and $P(t)$ is the pulse power's temporal profile. An example of the electron density evolution is presented in Fig.~\ref{Fig3}. One can see that the electron density in a narrow layer near the waveguide wall is almost an order of magnitude larger than near the axis. This difference increases rapidly with the neutral gas density growth.  
 
Note that the plasma density grows rapidly at the rear tail (late time) of the pulse, and the characteristic rise time of the plasma density is much smaller than the pulse duration. The importance of this will be discussed below.  

\section{Eigenmodes of a waveguide filled partially with plasma}

The neutral gas ionization process induced by a powerful microwave pulse described in Sect. III assumes that the structure of the microwave fields remains unchanged by the presence of plasma, which is correct for plasma density smaller than critical $n_{\rm crit}=\omega^2 m/4\pi e^2$. When the plasma density is close to critical, the structure of the wave's fields vary considerably. In order to trace this variation, let us consider the simple model of assuming a step-wise three-layer density distribution: 
\begin{equation}\label{eq1}
	n(r)=\left\{\begin{array}{clc}
		n_1,& r<r_1 &\mbox{(region I)},\\
		n_2,& r_1<r<r_2&\mbox{(region II)},\\
		n_3,& r_2<r<R&\mbox{(region III)}
	\end{array}
	\right.
\end{equation}
where $R$ is the waveguide radius. This plasma density profile is a rough description of the expected density distribution described in Sect. III. 

{A cold unmagnetized plasma can be considered as a medium with permittivity $\varepsilon(\omega)=1-\omega_p^2/\omega^2$, where $\omega_p$ is the electron Langmuir frequency, $\omega_p=\sqrt{4\pi e^2n_e/m}$, and $\omega$ is the electromagnetic wave frequency.  The axially-symmetric TM-mode in any region contains axial and radial electric fields, $E_z$ and $E_r$, and azimuthal magnetic field $H_\varphi$.
Matching the tangential components of the electromagnetic fields, $E_z$ and $H_\varphi$ at the interfaces between the different regions, one can find the spectrum $\omega(k)$ of the layered waveguide eigenmodes as the solution of a boundary value problem (see, e.g., \cite{Kinderdijk-1971}).} 

Even such a simple form of the plasma density distribution contains many parameters. Below, only such values of the parameters will be considered which correspond to the experimental conditions. In particular, because the plasma density in region II can exceed the ones in regions I and III, the model can be further simplified, by assuming $n_1=n_3=0$. 

\begin{figure}[h] 
	\centering \scalebox{0.45}{\includegraphics{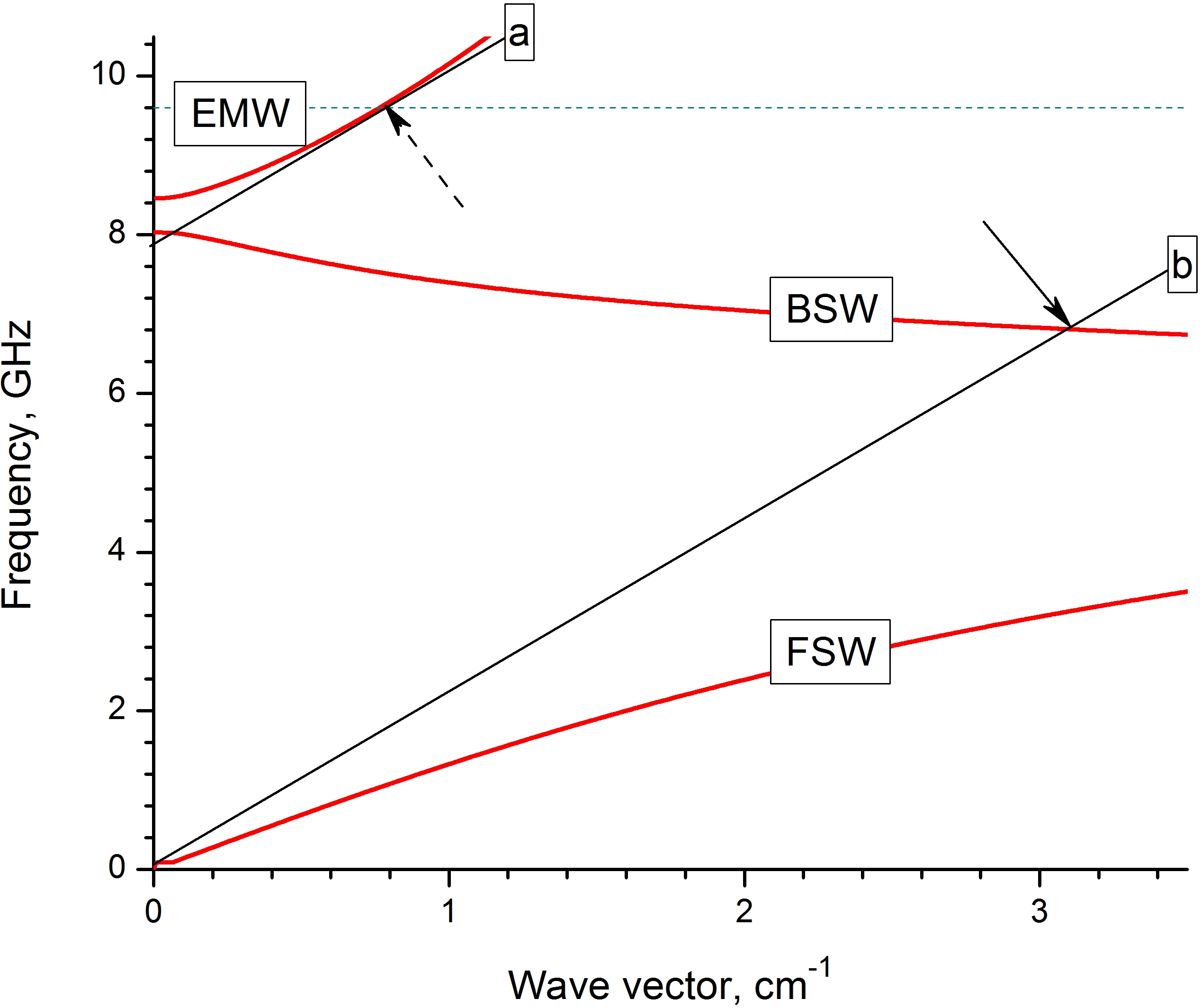}}
	\caption{Spectrum of the eigenmodes for $R=1.4$ cm, $r_1=1.0$ cm, $r_2=1.2$ cm, plasma density $n_e=8\cdot 10^{11}\,{\rm cm}^{-3}$.  Red thick lines -- volumetric electromagnetic wave (EMW), backward surface wave (BSW), and forward surface wave (FSW). The dashed horizontal line marks the experimental frequency. The dashed arrow points out the frequency $f$ and the wave vector $k$ of the incoming pulse, and line ``a'' is tangential to the dispersion curve at this point. Line ``b'' is parallel to line ``a''. The intersection between lime ``b'' and the dispersion curve BSW defines the frequency and wave vector of the excited backward surface wave.  }
	\label{Dispersion-1}
\end{figure}

\begin{figure}[h] 
	\centering \scalebox{0.4}{\includegraphics{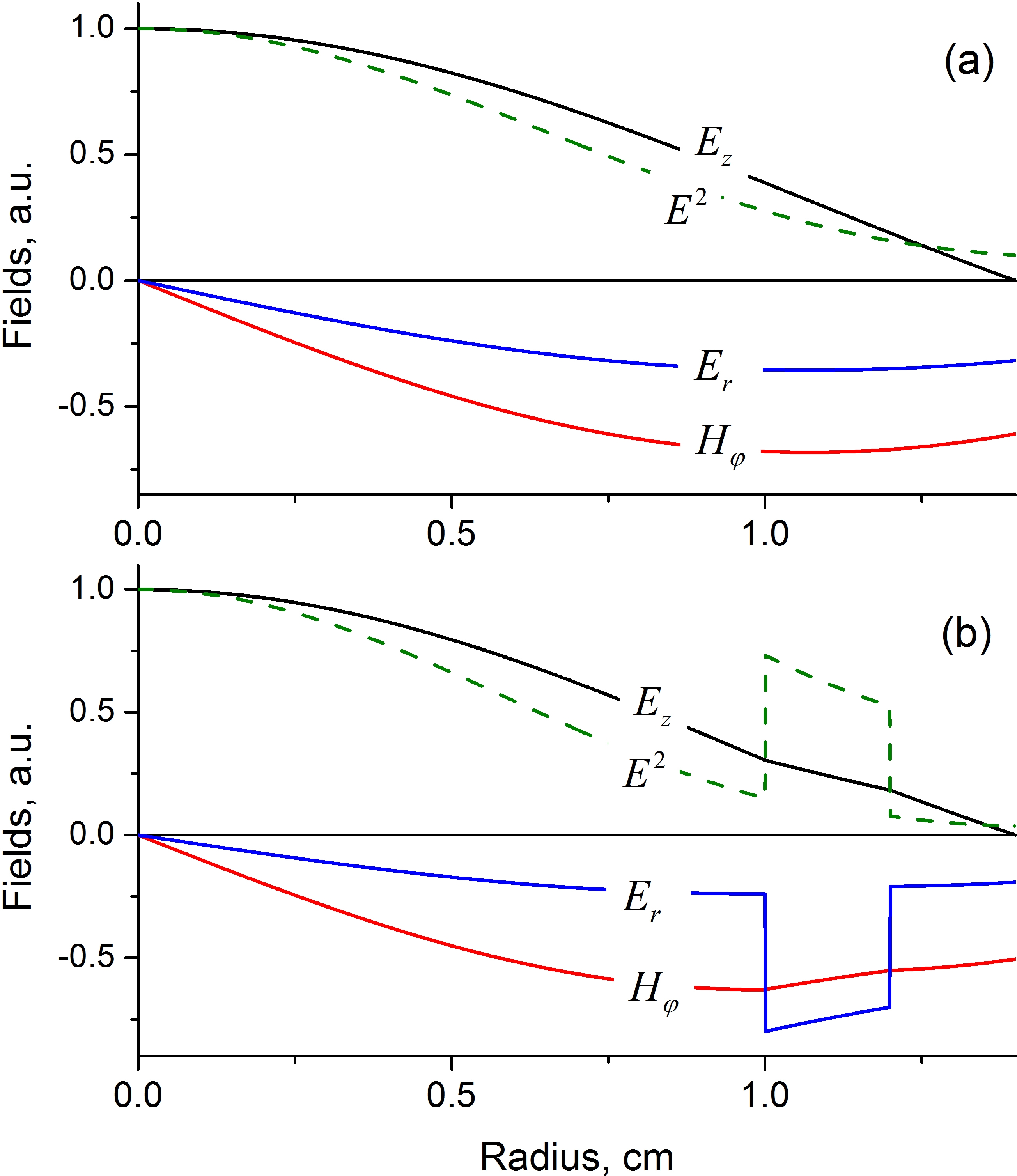}}
	\caption{{Radial structure of the electromagnetic wave fields (EMW mode in Fig.~\ref{Dispersion-1}, frequency 9.6 GHz). The radial dependence of the amplitudes of the electric and magnetic fields and the square of the total electric field in the empty waveguide (a) and in the presence of the plasma (b).} }
	\label{Fields-1}
\end{figure}

{The spectrum of eigenmodes of the waveguide used in the experiments, is shown in Fig.~\ref{Dispersion-1}.}
The presence of the plasma with a ring-shaped transversal profile  enriches the spectrum by two surface waves: a backward- and a forward-propagating mode. The spectrum of the electromagnetic wave changes only little if the plasma layer's thickness is small compared to the waveguide radius. On the contrary, the change in the structure of the wave's fields are very strong in Fig.~\ref{Fields-1}.

Compared to an empty waveguide, in the plasma-filled waveguide the amplitudes of the longitudinal component of the electric field, $E_z(r)$, and the azimuthal component of the magnetic field, $H_\phi(r)$, decrease over all cross sections, but not that much. In contrast, the variation in the radial electric field, $E_r(r)$, profile is considerably stronger. This component of the field increases strongly in region II, occupied by the plasma which means that the oscillatory energy of the plasma electrons $w\propto E_z^2+E_r^2$ exceeds the value calculated ignoring the change in the eigenmode fields. Because the ionization cross section is a decreasing function of the electron energy (in the energy range of interest), changes in the eigenmode fields structure suppresses ionization and the region with the most effective ionization is shifted closer to the waveguide axis, widening the plasma layer. {Note, that the step-like profile of the electric field in Fig>~\ref{Fields-1} reflects the step-like profile Eq.~(\ref{eq1})of the plasma density distribution. The real distribution is smooth, and the role of this choice will be discussed below.  }

In the numerical example relevant to Fig. 5, the plasma density was chosen so that the Langmuir frequency $\omega_p$ is equal to the empty waveguide cut-off frequency $\omega_c$. This is an arbitrary choice, but the analysis allows one to assume that the plasma density in the expanded layer is close to this value. Indeed, if the plasma density is small, so that $\omega_p\ll\omega_c$, the deformation of the eigenmode fields is small, and also the ionization in this region continues. When the density is large enough, so that $\omega_p\simeq\omega_c$, the ionization slows down. The density cannot be so large that $\omega_p\gg\omega_c$, because the waveguide becomes opaque for the incoming microwaves if the  frequency $\omega_{\rm EMW}$ is close to  the cut-off frequency, $\omega_c$, as is the case in the experiments.

Thus, a high-power microwave pulse, propagating in the waveguide filled with a neutral gas, can create a hollow, tubular plasma. Note that a similar hollow plasma column can be created in an unbounded system by a microwave pulse with a bell-shaped radial profile when the required power density can be achieved by focusing the microwave beam \cite{Shafir-2018, Cao-2018}, so that the hollow plasma is created in the vicinity of the focal plane only. A hollow plasma can also be created by a tubular electron beam, propagating in a neutral gas along a strong guiding magnetic field \cite{Kuzelev-1982}. Ionization of a neutral gas by a powerful tubular laser beam can also create a plasma with specified density distribution \cite{Laser}.

\section{Backward surface wave excitation and plasma glow striations}

Propagation of a powerful electromagnetic pulse in a preliminary prepared plasma is accompanied by the excitation of a wake, which is a slow eigenmode of either an unbounded plasma or a plasma-filled waveguide. Phase velocity of this eigenmode coincides with the group velocity of the pulse.{Charged  particles acceleration in the wake field, excited by a laser pulse in  plasma, is may be the best known example of this phenomenon (see, e.g., \cite{Esarey-2009}).  }It seems reasonable to assume that a similar wake is excited and remains in the plasma behind the pulse even though the plasma was created by the pulse itself. 

\begin{figure}[h] 
	\centering \scalebox{0.45}{\includegraphics{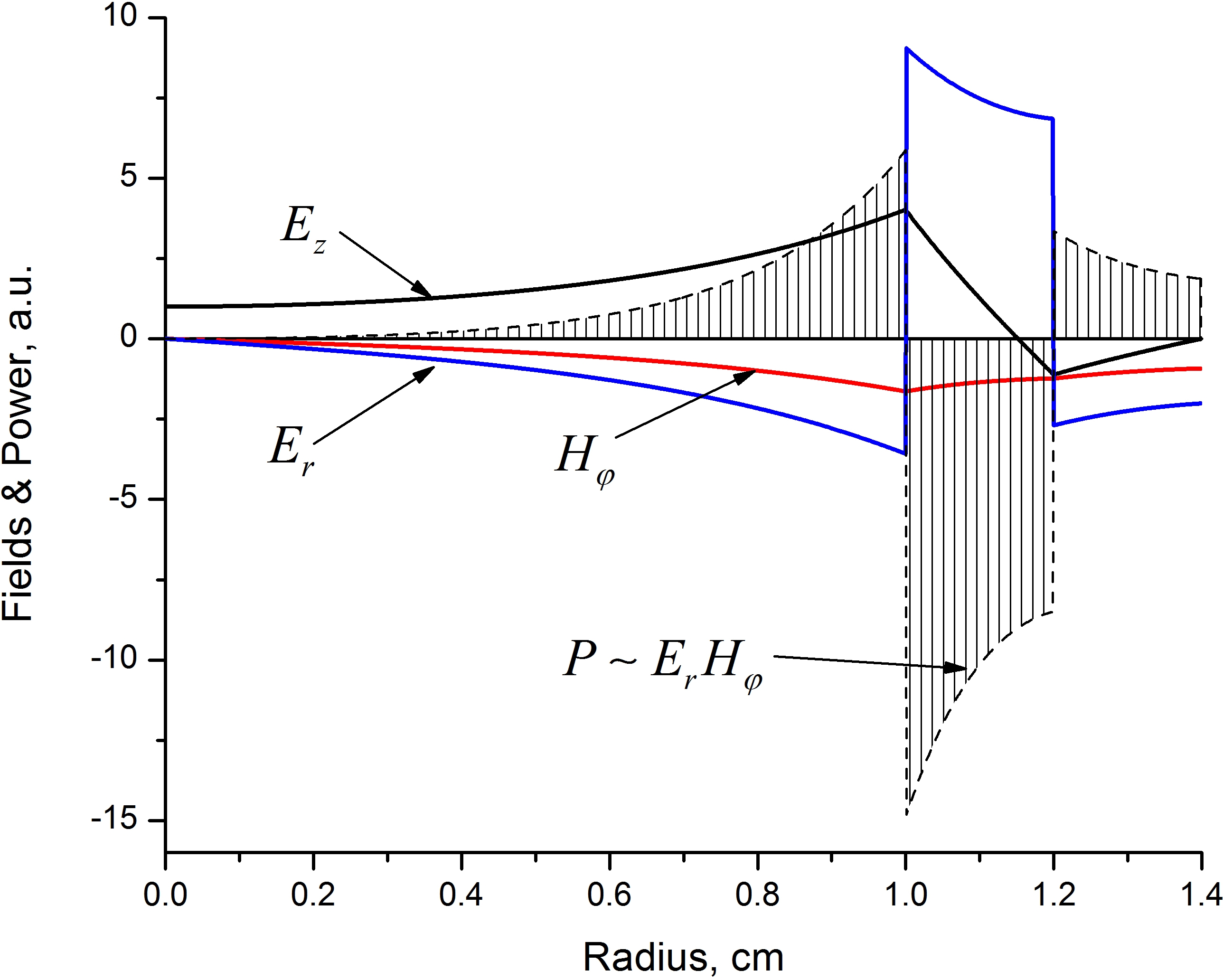}}
	\caption{Radial distribution of the electric ($E_r$, $E_z$), magnetic, ($H_\varphi$) fields, and longitudinal power flux density $P\sim E_rH_\varphi$ of the excited backward propagating surface wave. {The wave frequency and the wave vector are defined by resonance with the electromagnetic wave, as it is shown in Fig.~\ref{Dispersion-1}}}
	\label{BSW fields}
\end{figure}

Indeed, for the case considered, the perturbation, produced by an electromagnetic pulse, remains in a tubular plasma in the form of a backward surface wave. The wave number and the frequency of the exited surface wave are defined by \v{C}herenkov's resonance condition $\omega_{\rm BSW}(k)=k_{\rm BSW}v_g$, as shown in Fig.~\ref{Dispersion-1}. Here $\omega_{\rm BSW}$ and $k_{\rm BSW}$  are assigned to the backward surface wave and $v_g$ is the group velocity of the electromagnetic pulse. This wave can persist in the plasma for a long time after the pulse passage. Radial distribution of the electric and magnetic fields of this backward propagating surface wave is shown in Fig~\ref{BSW fields}. 

{Note  that the radial electric field and the energy flux  density $P$ change their polarity at the intersection of the plasma boundaries. This is a well-known property \cite{Nkoma-1974} of a surface wave propagating along the interface between two media with opposite signs of permittivity. The sign of the energy flux integrated over the waveguide cross section determines whether the wave is forward or backward propagating. }

Axially-symmetric surface waves with negative group velocity in waveguides containing a plasma column in a dielectric tube, separated from the conducting wall, were considered in \cite{Trivelpiece-1958}. Backward surface waves in a waveguide with a step-wise annular plasma column, similar to that considered in this article, were studied in \cite{Paik-1962}. It was shown here that the backward and forward propagating modes are associated with the inner and outer plasma surfaces, respectively. Surface waves in waveguides filled with smoothly distributed inhomogeneous plasma were studied in \cite{Kuzelev-2005,Kuzelev-2014}. It was shown that the absence of sharp boundaries between different layers leads to collision-less dissipation of the surface waves. The reason is the presence of resonant electrons in the smooth transition plasma layer.

Negative group velocity of the surface wave means that the wave, excited along the plasma-filled section of the waveguide, propagates in the opposite direction (upstream) towards the input window. Here the wave is reflected and propagates towards the downstream window. Because the phase velocity of the reflected wave is negative, these two incident and reflected waves form a \textit{standing} wave near the input window. Electrons, oscillating in the electric field of the standing wave antinodes, can excite and/or ionize molecules of the neutral gas, producing a periodically distributed plasma glow. 

It is worth noting that, under certain conditions, stationary striations in a glow discharge are the result of interference between two ionization waves \cite{Cooper-1958, Stirand-1967, Maruyama-1990}. Unlike backward-propagating surface waves, ionization waves are excited only in the discharge current  and disappear immediately after current interruption.    

As shown in the Sect. IV, the process of the plasma creation by the microwave pulse can be roughly divided into two phases. Initially, plasma is created in a narrow layer near the waveguide boundary. When the electron Langmuir frequency $\omega_p(r)$ in this layer approaches the cut-off frequency $\omega_c$, the plasma density growth slows down, and the second phase of the process begins. Now, ionization leads to widening of this layer with approximately constant plasma density. In the step-wise model, Eq.~(\ref{eq1}), it means that $n_2(t)\simeq {\rm const}$ and $r_1(t)$ decrease with time. Simultaneously, the cut-off frequency of the plasma-filled waveguide grows and, if the pulse duration is long enough, it reaches the incoming wave frequency $\omega_{\rm EMW}$ (which, in the experiment, is close to the cut-off frequency of the empty waveguide, $\omega_{\rm EMW}\simeq 1.17 \omega_c$) while the retained part of the pulse is reflected from the waveguide input window.   

\begin{figure}[h] 
	\centering \scalebox{0.4}{\includegraphics{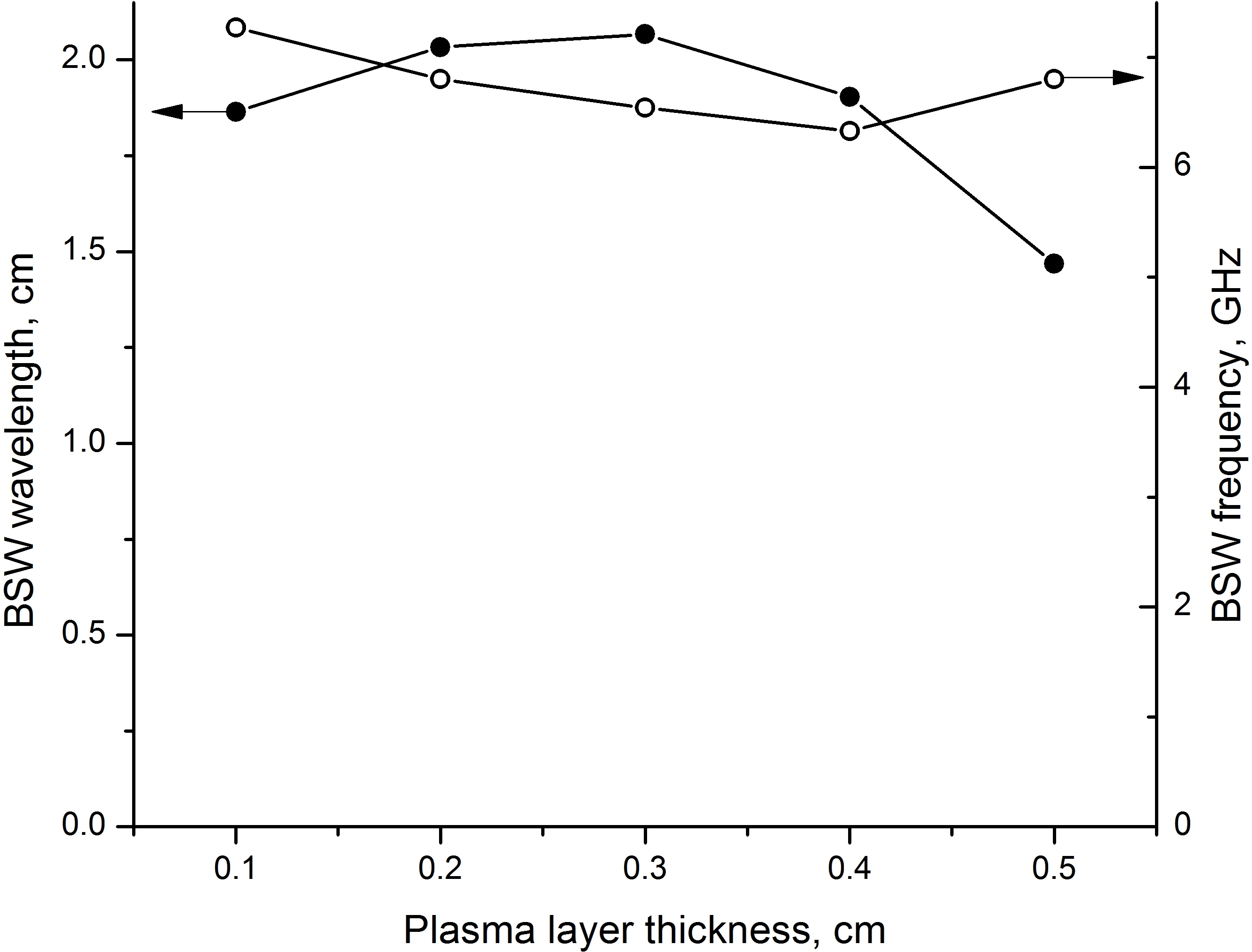}}
	\caption{Resonant wavelength (closed circles) and resonant frequency (closed circles) of the backward surface wave as functions of the plasma layer thickness. }
	\label{Wavelength}
\end{figure}

When the layer thickness $d=r_2-r_1$ varies with time, the pulse group velocity $v_g$ and the backward surface wave dispersion characteristics $\omega_{\rm BSW}(k_{\rm BSW})$ vary too. In this case, effective excitation of the surface wave is possible if the resonant frequency $\omega_{\rm BSW}$ and wave number $k_{\rm BSW}$ remain essentially unchanged during the pulse duration. In Fig.~\ref{Wavelength} the dependence of the resonant surface wave frequency and wavelength on the plasma layer thickness demonstrate relatively small dispersion. The characteristic wavelength is about 2 cm, so the distance between nodes of the standing wave is about 1 cm. Taking into account the roughness of the model, one can consider that this value agrees well with the experimentally observed period of the glow (~1.8 cm).  

\section{Discussion}

\begin{figure}[h] 
	\centering \scalebox{0.45}{\includegraphics{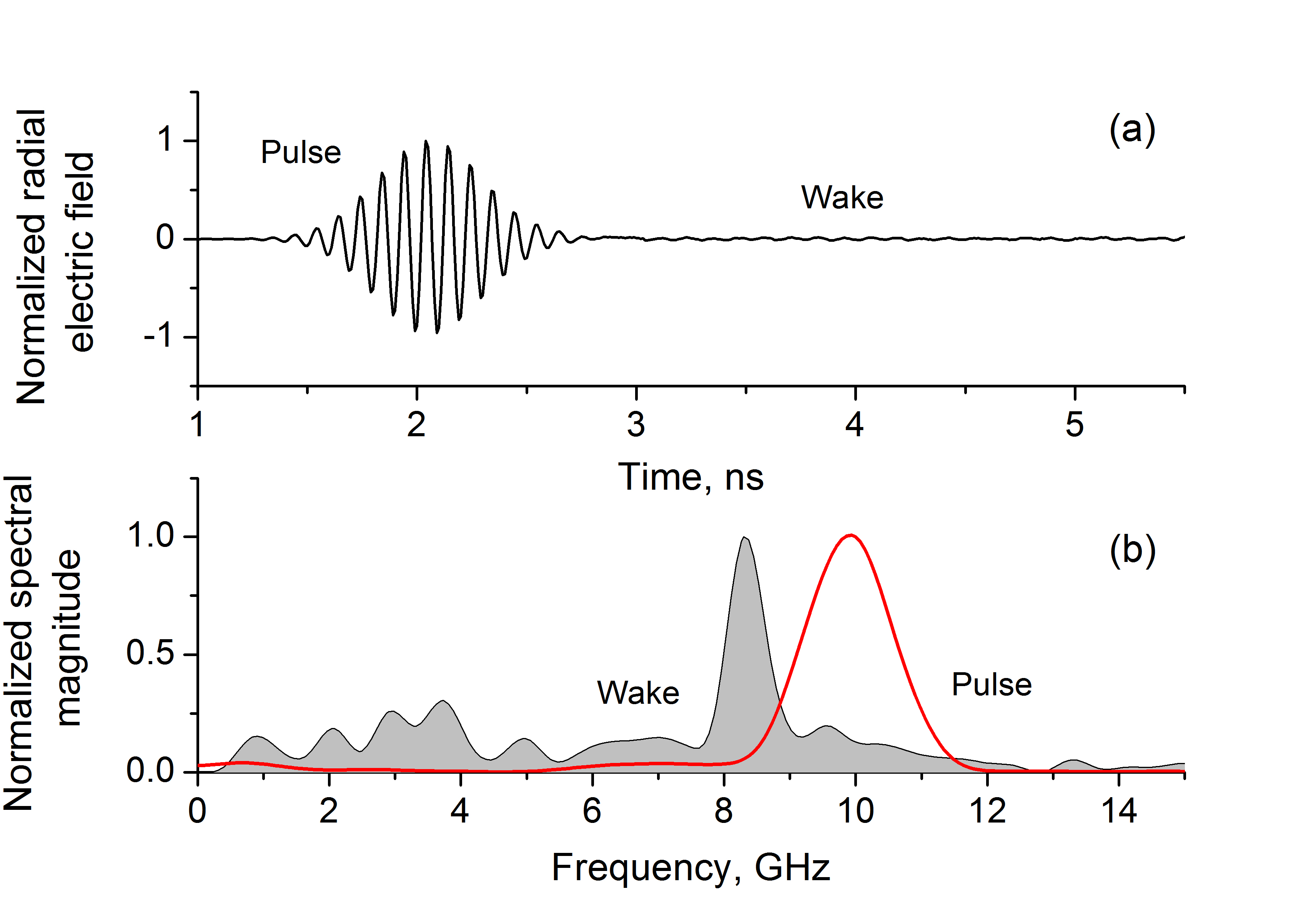}}
	\caption{a -- The time dependence of the radial electric field in the plasma layer. Long-living wake remains in plasma after the pulse passage. b -- Spectra of the electromagnetic pulse (red line) and the wake (filled grey). Note that the wake spectrum is narrower and shifted to lower frequencies. (See also Fig.~\ref{Dispersion-1}.   }
	\label{Field_Spectra}
\end{figure}

The resonant excitation of a wave of any nature by a moving perturbation is effective when the characteristic spatial scale of the perturbation is small compared with its wavelength, or otherwise, the amplitude of the wake behind the perturbation is small. As seen in Fig.~\ref{Dispersion-1}, the expected wavelength of the backward surface wave is smaller than the electromagnetic pulse wavelength and much smaller than the longitudinal dimension of the pulse. Thus, the electromagnetic pulse with parameters corresponding to the experimental conditions, moving through a tubular plasma, can excite a surface wave of a small amplitude. Numerical simulations using the 3D Particle in Cell (PIC) code Lsp (Large scale plasma) \cite{Welch-2001,Welch-2006}, confirmed this conclusion. Propagation of microwave pulse (9.6 GHz, 0.35 ns duration at FWHM, and 10 MW power) through cylindrical waveguide of radius 1.4 cm, part of which is filled by the tubular plasma (inner and outer radii are 1.0 and 1.2 cm, respectively) of density $8\cdot 10^{11}$ approximately modeled the experimental conditions. Small microwave power correspond to the end of the pulse, when the plasma density reaches its maximum.   Results of simulation, presented in Fig.~\ref{Field_Spectra}, confirm, that the surface wave of small amplitude is indeed excited behind the microwave pulse.
The main peak of the wake spectrum, depicted in Fig.~\ref{Field_Spectra}, is located in the region of 8-9 GHz, that is in good agreement with Fig.~\ref{Wavelength}. Note that the spectral width of the wake is smaller than that of the pulse because of the following reason. A short pulse (such as that considered) is characterized by a wide spread of group velocity values rather than a single velocity. Consequently, the resonant condition $\omega_{\rm BSW}(k_{\rm BSW})=k_{\rm BSW}v_g$ is satisfied for a set of surface waves, but, due to the weak dependence $\omega_{\rm BSW}(k_{\rm BSW})$ (see Fig.~\ref{Dispersion-1}), the spectrum of the excited surface waves remains narrow. This property of the wake spectrum and its position strongly supports the backward flowing surface wave description.    

Despite that the excitation of the backward flowing surface wave by the electromagnetic pulse is confirmed in numerical simulations, the assumption that this wave is responsible for the observed spatially-periodic plasma glow casts doubt, because of the small amplitude of the surface wave. This discrepancy can be resolved in the following way. 

In the performed numerical simulation it was assumed that the plasma density is stationary and homogeneously distributed along the waveguide. However, in the experiment the plasma is created by the propagating electromagnetic pulse itself. Then, the temporal (and longitudinal) profile of plasma density, which is shown in Fig.~\ref{Fig3}, looks like shock wave and propagates along the waveguide together with the electromagnetic pulse. In other words, the surface wave propagates in the media with time-dependent, traveling parameters. The characteristic spatial scale of the plasma density variation is much smaller than the pulse length and can be of the order of the surface wave wavelength. The amplitude of the resonant wave, excited by such abrupt traveling variation of the medium parameters, is much larger compared to the case, when the medium parameters are varied smoothly \cite{Kravtsov-1990,Stepanov-1993}. 

The possible source of plasma perturbation, which remains behind the microwave pulse and propagates in the form of surface wave, is the ponderomotive force, $F_p$, which pushes out electrons from the region with enhanced amplitude of microwave electric field. The polarization electric field $E_p$ between immobile ions and shifted from their equilibrium position electrons remains in plasma, if the ponderomotive force of the pulse rear tail decreases rather fast with time, during several periods of microwave oscillations. 
This field and disturbed electron density evolve then as the surface wave, whose amplitude can be roughly estimated in the following manner. 

%%%%%%%%%%%%%%%%%%%%%%%%%%%%%%%%%%%%%%%%%%%%%%%%%%%

The equality of the ponderomotive, $F_p$, and Lorentz, $eE_p$, forces defines the polarization electric field amplitude $E_p$, which is the surface wave amplitude:
\begin{equation}
	\label{eq4}
	eE_p=mc^2\nabla(a^2/2),
\end{equation}
where $a=eE_r(r_p)/mc\omega$ and $E_r(r_p)$ is the amplitude of the radial electric field of the electromagnetic wave in the plasma layer. All the calculations below will be carried out for the same values of parameters, which were used   
to determine the dispersion relations and the wave fields, depicted in Figs.~\ref{Dispersion-1}, \ref{Fields-1}, and \ref{BSW fields}: $r_1=1.0$ cm, $r_2=1.2$ cm, $r_p=1.1$ cm, plasma density $n_e=8\cdot 10^{11}$ cm$^{-3}$.

Using the known structure of the wave fields (see Fig.~\ref{Fields-1}), one can define the value of parameter $a$, which corresponds to a given value of the microwave pulse power $P_{pulse}$ as $a(P_{pulse})$. For a typical value of the pulse peak power $P_{pulse}\simeq250$ MW, $a\simeq 0.38$. 

Next, the value of the radial gradient $\nabla(a^2)$ can be estimated as $a^2/d$, where $d=0.2$ cm is the thickness of the plasma layer. Using the fields structure of the backward surface wave (see Fig.~\ref{BSW fields}) and the value of the field $E_p$, one can calculate the power $P_{BSW}$ of this wave: $P_{BSW}\simeq 0.14 P_{pulse}\simeq 41$ MW. Here it is necessary to note that these estimations were performed under the above discussed assumption that the plasma density is close to its cut-off value.  It was also noted earlier, that the plasma density approaches this value at the rear tail of the microwave pulse  (see Fig.~\ref{Fig3}), where the pulse power is much smaller than its maximal value, that is, only a few percent of the peak power pulse $P_{pulse}$. Consequently, the backward surface wave power is approximately two-three orders of magnitude smaller than that calculated above. Thus, the power of the backward propagating surface wave does not exceed a few hundred kilowatts. For example, when the power of the rear part of the microwave pulse, which is responsible for the surface wave excitation, is only 5\% from the peak power,  $P_{BSW}\simeq 100$ kW. The radial electric field of the backward propagating surface wave reaches 7 kV/cm, and the energy of the electrons, oscillating in this field, reaches 400 V, sufficient for the excitation and ionization of the neutral gas. Thus, the stationary periodic plasma glow can indeed be associated with the excitation of the backward surface wave.

\section{Conclusions}

It was observed experimentally that ionization of a neutral gas by a sub-nanosecond microwave pulse of hundreds of MW power propagating in a waveguide is accompanied by a new phenomenon. Namely, spatially-periodic stratification of the plasma glow appears near the waveguide wall 5 ns after the microwave pulse’s passage and continues during ~70 ns. In order to explain this phenomenon, a theoretical model was developed. Analytical and numerical studies carried out in the frame of this model, indicate that such a powerful microwave pulse   
produces a tubular plasma, with the density mainly concentrated in a thin layer near the waveguide wall. A surface wave with negative group velocity can be excited in such plasma-filled waveguide. The wake in the form of this backward propagating surface wave remains in the waveguide long after the pulse's passage. A backward propagating surface wave is reflected from the input window and in combination with the incident wave forms a standing wave. Plasma electrons, oscillating in the electric field of this standing wave's antinodes, excite and/or ionize molecules of the neutral gas, producing the experimentally observed periodically distributed plasma glow along the waveguide.   

This scenario is based on a theoretical model supported by numerical simulations and analysis. We cannot present indisputable experimental facts that the observed spatially-periodic plasma glow appeared exactly in that way as it is described in the paper. Should this be the case then this phenomenon would be the first non-direct observation of the backward surface wave, propagating along a tubular plasma in the absence of magnetic field. In future experiments we will attempt to confirm the mechanism behind this interesting phenomenon. 
   
\section*{Acknowledgements}
 
The authors are grateful to E. Flyat for his technical support. This study was supported by Pazy Foundation Grant No. 2032056.

\end{document}